\documentstyle[preprint,aps]{revtex}
\newcommand{\gra}[1]{\mbox{\boldmath $ #1 $}}

\newcommand{\bfigc}{\begin{figure}\begin{center}}
\newcommand{\efigc}{\end{center}\end{figure}}
\def\eg{{\it e.g.}\ }  \def\ie{{\it i.e.}\ }

\input{epsf}

\begin{document}

\title{Analysis of cancellation in two-dimensional magnetohydrodynamic turbulence}

\author{L. Sorriso--Valvo$^{1,2}$, V. Carbone$^1$, A. Noullez$^2$,
H. Politano$^2$, A. Pouquet$^3$ and P. Veltri$^1$}

\address{$^1$ Dipartimento di Fisica and Istituto di Fisica della Materia \\
Universit\`a della Calabria, Rende (CS), Italy}
\address{$^2$ Lab. Cassini, Observatoire de la Cote d'Azur, Nice (France)}
\address{$^3$ ASP/NCAR, PO Box 3000, Boulder, CO, USA}

\maketitle
\date{\today}
\begin{abstract}

A signed measure analysis of two-dimensional
intermittent magnetohydrodynamic turbulence is presented.
This kind of analysis is performed to characterize the scaling behavior of
the sign-oscillating flow structures, and their geometrical properties.
In particular, it is observed that cancellations between positive and negative
contributions of the field inside structures, are inhibited
for scales smaller than the Taylor microscale, and stop near the dissipative
scale. Moreover, from a simple geometrical argument, the relationship 
between the cancellation exponent and the typical fractal
dimension of the structures in the flow is obtained.
\vskip 1cm
\end{abstract}

\newpage
\section{Introduction}
\label{sec1}
Self-similarity is a signature of complex flows with strong nonlinearities.
Huge amount of efforts have been deployed in determining scaling laws
of energy spectra and, more generally speaking, of structure functions 
(see \eg Ref.~1 and~2), both in geophysical and astrophysical flows, in
the laboratory and using numerical simulations as well.
Non-linear behavior of the scaling exponents is generally interpreted as due
to the presence of strong localized structures both in space and time.
Indeed, structures have been observed in turbulent flows for a long time:
sheets, spirals and filaments of vorticity both in the 
incompressible~\cite{siggia81,kerr,shejackson,vincentmeneguzzi91} 
and compressible case~\cite{porter}, in the latter with shocks as well. In 
magnetohydrodynamics (MHD), current and vorticity sheets seem to prevail in 
three dimensions~\cite{nordlund,brandenburg,politano95,galsgood,kerrmhd,bisk3d} 
whereas in two-dimensional space, current and vorticity filaments are 
the main identifiable structures~\cite{biskamp89,politano89,grauermarliani}. 
Such features are locally complex, with rapid fluctuations
corresponding to flux cancellations, in three dimensions~\cite{politano95} 
as well as in two dimensions, and as first strikingly observed in chaotic 
dynamos using mappings~\cite{ott}; in such cases, very high Reynolds number 
flows can be modeled and cancellations of flux occur at all scales, 
leading to a very complex magnetic field. 

We study in this paper the geometrical properties of the vorticity 
and current fields (as well as their combination through the Els\"asser 
variables defined below). The data stems from a direct numerical simulation 
in two space dimensions and our study focuses on the computation of
the cancellation exponent introduced by Ott {\it et al.}~\cite{ott}. 
These exponents allow for a simple characterization of the flows 
and can be linked to the fractal dimension of the typical structures.

The present paper is organized as follows.
In Section~\ref{sec2}, we recall the definitions of the signed measure and
cancellation exponent. In Section~\ref{sec3}, we describe the data 
stemming from a numerical simulation, and the main features of the fields are 
characterized by the typical scales of the flow.
Section~\ref{sec4} describes 
the results obtained from the analysis of the cancellation properties of our 
flow.
In Section~\ref{sec5}, we introduce a simple model, based on geometrical
arguments, which connect the cancellation exponent to the fractal dimension 
of the structures. Our results are discussed and the main conclusions of the
paper are re-emphasized in Section~\ref{sec6}.

\section{The signed measure and the cancellation exponent}
\label{sec2}

Intermittency in turbulent fields is related to the presence of structures at
different scales. The proximity of such structures may lead to rapid changes 
of sign in the fields derivatives. 
In order to describe the scaling properties of sign oscillations,
Ott~{\it et al.}~\cite{ott} introduced the concept of {\it sign-singular 
measures}. 
In analogy to probability measures (positive definite), 
the signed measure of a zero-mean scalar field $f(\gra r)$, defined on 
a set $Q(L)$ of size $L$, can be introduced in the $d$-dimensional case. 
Let $\{Q_i(l)\} \subset Q(L)$ be a hierarchy of disjoint subsets of size $l$ 
covering $Q(L)$. Then, for each scale~$l$ and for each set of boxes~$Q_i(l)$,  
the signed measure is defined as
 \begin{equation}
 \mu_i(l)=\frac{\int_{Q_i(l)}\;d\gra{r}\;f(\gra r)}
 {\int_{Q(L)}\;d\gra{r}\;|f(\gra r)|} \ .
 \label{mu}
 \end{equation}
The choice of the normalization factor (the denominator in Eq. (\ref{mu})) 
will be justified below.
The signed measure thus can be interpreted as the difference between 
two probability measures, one for the positive part and the other for the 
negative part of the field~\cite{halmos}.

As the scale of the subset~$Q_i(l)$ increases, cancellations between opposite
sign structures, and then between the two probability measures, come into play. 
This feature can be characterized by investigating the scaling properties 
of the partition function\,:
 \begin{equation}
 \chi(l) =\sum_{Q_i(l)} |\mu_i(l)|
 \label{chi}
 \end{equation}
where the sum is extended to all disjoint subset~$Q_i(l)$. It is clear that 
if~$\mu_i(l)$ is a probability measure, as well as if the field is defined 
in sign, the choice of the normalization factor in (\ref{mu})
leads to $\chi(l)=1$. On the other hand, if cancellations are
present, we can expect that their effect is less and less important as the
scale~$l$ decreases, so that $\chi(l)$ increases. 
In this latter case, we can investigate the scaling behaviors of the weakness of
cancellations by a scaling exponent $\kappa$, defined through 
$\chi(l)\sim l^{-\kappa}$, where $\kappa$ is called the 
{\it cancellation exponent}. 
This exponent represents a quantitative measure of the cancellation efficiency.
Furthermore, if this scaling law exists, in analogy with the singularities of 
probability measures, the signed measure is called sign-singular.
If the field is smooth, then trivially $\kappa=0$. It can be shown that for
homogeneous fields with discontinuities, $\kappa=1$,
whereas for stochastic processes $\kappa=d/2$~\cite{vains2},
$d$~being the dimensionality of the space.
More generally, as will be shown in section \ref{sec6}, 
if the field~$\gra f(\gra r)$ is homogeneous with a H\"older 
scaling exponent~$h$, that is if $\langle \|\Delta \gra f(\gra l)\| 
\rangle = \langle \|\gra f(\gra r+\gra l)-\gra f(\gra r)\| 
\rangle \sim l^h$, then $\kappa=-h$~\cite{vains2,chhabra}.
This relation between the cancellation exponent and scaling exponents can be
generalized to higher order partition functions and structure functions 
$S_p(l) = \langle \Delta^\parallel \gra f(l)^p \rangle \sim l^{\zeta_p}$ 
(where $\Delta^\parallel \gra f(\gra{l}) = [\gra f(\gra{r}+\gra{l})-\gra 
f(\gra{r})]\cdot \hat{\gra l}$ defines the longitudinal
increments of~$\gra f$ on scale~$l$) to determine the generalized dimensions 
of the set where the singularities of the field~$f$ live~\cite{vains2,vains1}.

Besides the work of Ott~{\it et al.}~\cite{ott} on fast magnetic dynamos, 
several flows in the context of plasma turbulence have been examined
using such concepts, in order
to characterize the oscillating features of turbulent fields. 
For example, the magnetic helicity in the solar wind has 
been shown to be sign-singular~\cite{carbruno}, and solar photospheric 
velocity patterns~\cite{consolini} as well. Finally, the
variations of the cancellation exponent of the current helicity between 
the pre-- and post--flare periods for major
H-$\alpha$ solar flares, have been used to evidence variations in the
heliospheric magnetic field structures of active regions~\cite{abram}.

\section{The numerical simulation and the typical scales of the flow}
\label{sec3}

The data we analyze with the technique described in the previous
section stems from a numerical simulation of the two-dimensional ($d=2$)
magnetohydrodynamics (MHD) equations written below in terms of the Els\"asser
variables~$\gra{z^\pm}=\gra{v} \pm \gra{b}$

\begin{equation}
(\partial_t+\gra{z^{\mp}}\cdot \nabla)\gra{z^{\pm}} = -\nabla P_* +
\nu^{\pm}\nabla^2\gra{z^{\pm}} + \nu^{\mp}\nabla^2\gra{z^{\mp}} +\gra{F^{\pm}}
\label{mhd}
\end{equation}
where $P_*$~is the total pressure, $2\nu^{\pm}=\nu \pm \eta$, $\nu$~being the
viscosity and $\eta$~the magnetic diffusivity, while $\gra{F^{\pm}}$ are
the forcing terms. The incompressibility conditions 
$\nabla \cdot \gra{z^{\pm}} = 0$ complete the equations.

The MHD equations have nonlinear scaling properties 
so that Kolmogorov-like arguments~\cite{k41} can be performed to evaluate 
dimensionally the scaling properties of structure functions, but taking into 
account the specificity of MHD flows. 
The less efficient interaction rate between the~$\gra z^{\pm}$ structures may 
well lead to scaling laws that differ from the classical Kolmogorov case, as
first derived in the so--called Iroshnikov-Kraichnan 
phenomenology~(IK)~\cite{ik} for globally isotropic flows ({\it i.e.} in the
absence of a strong uniform magnetic field ${\bf B}_0$ at zero frequency), 
a phenomenology which takes into account the presence of Alfv\'enic 
fluctuations in the plasma. 
Small-scale intermittency is present in MHD as in the fluid case,
and is related to the presence of coherent structures which locally in space 
obviously break the isotropy assumption. However, in the Kolmogorov spirit, 
isotropy is recovered on average. The effects of these
structures on the phenomenology for MHD flows 
have been pointed out and discussed in several
ways (see, \eg Ref.~10 and~27).

In the present paper, we want to analyze the oscillating character 
of such structures, and their geometrical properties. 
Equations~(\ref{mhd}) are solved with a pseudo-spectral method on a
grid of~$1024^2$ points, with $2\pi$-periodic boundary conditions. The forcing
terms $\gra{F^{\pm}}$ maintain constant the Fourier modes $\gra{k}$ with 
$|\gra{k}|=1$. The magnetic Prandtl number $P_r=\nu/\eta$ is unity, so that
$\nu^+=\nu=\eta=8\times 10^{-4}$. The correlations between the velocity 
and the magnetic fields leads initially to $\rho_C = 
2\langle\gra v\cdot\gra b\rangle/\langle|\gra v|^2+|\gra b|^2\rangle \sim 6\%$.
The time-averaged Reynolds number is $R_e \sim 1600$ and the integral scale is 
$l_0/L=0.25\pm 0.02$, where $L=2\pi$ is the size of the computational box. 
Further details about the simulation can be found in Ref.~28 and~29. 
In the present work, we analyze ten samples separated by approximately
16~eddy turnover times once all transients have died out and a 
statistically steady state has been reached.

A snapshot of the current is shown in Figure~\ref{fig1}. The top picture shows 
the local accumulation of sheets, probably through folding, whereas the 
bottom picture clearly reveals the oscillating character of the flow structures.
The analysis of the probability distribution functions (PDFs) of the field 
fluctuations~\cite{pdfs}, and of the scaling exponents of the
structure functions~\cite{struct} has revealed that both the Els\"asser fields, 
as well as the magnetic field and the velocity, are strongly intermittent, as
already visualized in earlier computations~\cite{politano89,biskamp89}.
This intermittency is already quite evident when one computes the flatness 
factor $F(s)=\langle[\Delta^\parallel \gra f(s)]^4\rangle
/\langle[\Delta^\parallel \gra f(s)]^2\rangle^2$ that goes from the gaussian
value~3 at large scales $s$ for all fields, to values 
$F\sim 40$ for $\gra f \equiv \gra v$, $F\sim 80$ for $\gra f \equiv \gra b$, 
$F\sim 65$ for $\gra f \equiv \gra{z^\pm}$ at the smallest resolved 
scale.
This dramatic change in flatness reflects the change in shape of the PDFs 
through the scales, as quantified in Ref.~29. 

Following Politano and Pouquet~\cite{teorema}, it can be shown that the third 
order correlators defined below obey, in the inertial range, the following 
scaling law in dimension~$d$
 \begin{equation}
 Y_3^{\pm}(l) = \langle \Delta^\parallel \gra z^{\mp}(l) \;
                      \|\Delta \gra z^{\pm}(l)\|^2 \rangle
              = -\frac{4}{d}\;\epsilon^{\pm}\;l
 \label{teorema}
 \end{equation}  
where $\Delta \gra z^{\pm}(l)=\gra z^{\pm}(\gra{r+l})-\gra z^{\pm}(\gra r)$,
is the vector increment of the field $\gra z^{\pm}$, and $\Delta^\parallel$ is
its (scalar) longitudinal projection; thus, the flux functions 
$Y_3^{\pm}(l)$ involve all components of the physical fields. 
The coefficient~$\epsilon^{\pm}$ are the 
mean energy transfer rates of the $\gra{z^\pm}$ variables. 
We recall that in the framework of the Kolmogorov phenomenology, the mean 
energy transfer rate is assumed to be equal to the mean energy injection 
rate, as well as to the mean energy dissipation rate. 
In order to estimate~$\epsilon^{\pm}$, the data are fitted for each time in
the ranges where the relation~(\ref{teorema}) is verified. Averaging in time 
in order to decrease statistical errors, we evaluate $\epsilon^+ \sim 550$ and 
$\epsilon^- \sim 440$. The difference between the two $\pm$ rates
is linked to the correlation between the velocity and the magnetic field, and
should increase with increasing~$\rho_C$ and disappear only in the limit 
of vanishing $\rho_C$. Figure~\ref{fig2} displays the $Y_3^{\pm}(l)$  
third order correlators. It appears that the exact scaling law (\ref{teorema}) 
-- obtained in the limit of very large Reynolds numbers, otherwise correction 
terms not written here would arise due to dissipation -- 
is better verified in the case of $Y_3^+$ than for $Y_3^-$,
where the scaling range coincide 
with the inertial range as determined in Ref.~28 (see below).
Note that, because the fluxes are odd-order in the fields, cancellations
occur that render scaling laws difficult to unravel. For that reason, it is
customary to estimate the inertial range 
by computing the modified third order correlators, taking absolute values to
obtain $L_3^{\pm}(l)=\langle |\Delta^\parallel \gra z^{\mp}(l)| \;
\|\Delta \gra z^{\pm}(l)\|^2 \rangle \sim l$.
This leads to an inertial range extending from~$0.01$ to~$0.1$ in $l/L$  
for both correlators~\cite{struct}.
The difference in scaling quality between the plus and minus third order 
correlators, or higher order moments of the $\gra{z^+}$ and $\gra{z^-}$ fields,
has already been observed, even when using absolute moments to reduce 
statistical errors~\cite{bisk3d,gomez,struct}. 
This asymmetry between $\gra{z^+}$ and $\gra{z^-}$ could be due to a significant 
role of the velocity-magnetic field correlations on the dynamics of MHD flows,
even when $\rho_C\sim 0$. Indeed, the local $\gra v\cdot \gra b$, not
positive definite, is known to develop strong fluctuations (see, \eg Ref. 31), 
precluding identical behavior for $\gra{z^\pm}$ fields.

We can use the values of~$\epsilon^{\pm}$ computed from 
the law~(\ref{teorema}) to estimate -- in the framework of the IK 
phenomenology~\cite{ik} -- the typical scales of the flow, \ie
the dissipative scales $l_d^{\pm}=\left(\nu^2B_0/\epsilon^{\pm}\right)^{1/3}$ 
and the Taylor microscales 
$\lambda^{\pm}=\left(2\nu E^{\pm}/\epsilon^{\pm}\right)^{1/2}$. 
Here~$B_0$ represents the {\sl r.m.s.} large--scale magnetic field, 
and~$E^{\pm}$ are the energies associated with the Els\"asser fields, which are
the ideal invariants of the flow, together with the~$\langle a^2
\rangle$~correlation, where $\gra{b}=\nabla \times \gra{a}$
with $\gra{a}$ the magnetic potential.
The computed scales we obtain are given in Table~\ref{table}, 
and indicated in the orizontal axis of Figure~\ref{fig2} as well.
In MHD turbulence, it is not possible to unambiguously define directly
from the dissipation rates, the 
Taylor and the dissipative scales for the magnetic and velocity fields 
independently. However, using several assumptions, namely 
(i) equipartition, \ie 
$\langle \gra{v^2} \rangle/2 \sim \langle \gra{b^2} \rangle/2$ for the 
kinetic and magnetic energies, 
(ii) weak total correlation, 
$\langle \gra{v \cdot b} \rangle \sim 0$, and 
(iii) unit magnetic Prandtl number, $P_r \sim 1$, one can then
define a dissipation scale 
$l_d=\left(\nu^2B_0/\epsilon\right)^{1/3}$ and a Taylor microscale 
$\lambda=\left(2\nu E/\epsilon\right)^{1/2}$, where $E$ stands for the total
energy, and $\epsilon$ for its mean transfer and dissipation rates. Doing so,
and noting that $\epsilon=(\epsilon^+ + \epsilon^-)/2$, we obtain 
$l_d/L=(2.2\pm 0.4)\times 10^{-3}$ and $\lambda/L=1.5\pm 0.4\times 10^{-2}$.
These scales are very close to the ones computed for the Els\"asser fields. We
are thus left to use $l_d^{\pm}$ and  $\lambda^{\pm}$ as the typical scales of
the flow to refer to.

\section{Scaling properties of the signed measure in the numerical data}
\label{sec4}

We now analyze the scaling properties of the signed measure for the numerical 
data described above. As in the previous section for the evaluation of the 
energy transfer rates, we do so separately for
the different snapshots of the simulation, and we report
only the computed averages, while the error bars give an estimate of
the dispersion of such values.

We first define the signed scalar densities we need from the
physical variables $\gra v$, $\gra b$ and $\gra z^{\pm}$.
For two-dimensional MHD flows, it is convenient to use the rotational of
these variables.
Indeed, using the Stokes theorem, the circulation of the magnetic 
field along any closed contour, is equal to the 
normal flux of the current crossing the corresponding surface $Q_i(l)$ 
$\oint_{C_i(l)} \gra b(\gra r) \cdot d\gra \ell = \int_{Q_i(l)}
 \gra J(\gra r) \cdot \hat{\gra n}\; d\sigma \; .$
The sign of the current density flux is determined by the clockwise or
counterclockwise circulation of the magnetic field. In a two-dimensional
geometry, the magnetic field lies in the $(x,y)$~plane, namely 
${\gra b}(x,y) = (b_x,b_y,0)$, and the current density 
${\gra J}(x,y) = {\gra \nabla} \times {\gra b}(x,y)$ is
along the $z$-direction, perpendicular to the plane.
The normal current flux density is then simply 
$\gra J \cdot \hat{\gra n}\,d\sigma = J_z(x,y)\,dx\,dy$, 
so that $J_z(x,y) \equiv J(x,y)$ can be used as our scalar density. 
The same argument holds for the vorticity $\gra{\omega}$ and for 
$\gra{\omega^{\pm}=\omega \pm j}$, the rotationals of the Els\"asser fields. 
We show in Figure~\ref{fig3} the coarse-grained
current density flux for four different values of the box size $l$,
at one particular time ($t=7.3 \sim 45\,\tau_{NL}$, where $\tau_{NL}$ is the eddy
turnover time once the statistically steady state is reached). 
As can be seen, strong oscillating signed structures appear at all scales, 
and cancellation effects clearly increase with the box size.
In Figure~\ref{fig4} we present the function $\chi^J(l)$, computed 
in the case of the current $J(x,y)$ at each time and averaged over the ten 
temporal fields we analyze. 
A clear scaling range is visible, so that a cancellation 
exponent $\kappa^J=0.43$ can be obtained by performing a least-square fit 
of the type 
 \begin{equation}
 \chi(l)=C\left(\frac{l}{L}\right)^{-\kappa} \ .
 \label{powlaw}
 \end{equation}
We now try to relate the scales which characterize the cancellations to the 
typical scales of the flow itself, namely the Taylor microscales $\lambda^{\pm}$ 
and the dissipative ones $l_d^{\pm}$. 
The scaling range is determined by looking at the local slope of 
$\chi^J(l)$ (not shown), and extends from 
$l^J_{\star}=0.015L$ to $l^J_u=0.12L$. 
As can be observed, the range in which the scaling (\ref{powlaw}) holds
is mostly embedded within the inertial range of the flow, 
given in section \ref{sec3}.
For scales larger than $l_u$, the partition function still decays as a 
power-law, but with a larger 
exponent, approaching the typical behavior expected for a completely
uncorrelated point field, with $\kappa=d/2=1$,
but the poor statistics at such large scales prevent us from
discussing that behavior any further.
For scales smaller than $l_{\star}$, the partition function 
slowly saturates at the asymptotic value $\chi(l)=1$. 
The scale~$l_S$, at which the saturation can be considered achieved,  
is computed as the intersection scale between 
the power-law (\ref{powlaw}) and the limit value $\chi(l)=1$, giving 
$l^J_S=0.005L$ for the current. 
It is worth pointing out that, due to the choice of normalization in the signed 
measure, for any dataset with a finite resolution~$l_{\rm min}$, 
the partition function will trivially go to unity at that scale. 
The fact that, in our case, saturation of $\chi(l)$ persists for scales larger
than $l_{\rm min}$,
demonstrates that the field is well resolved and thus regular 
at the resolution of the numerical simulation.
The departure from a constant behavior of~$\chi(l)$ (with $\kappa=0$), 
occurs only at some yet larger scale, which can be identified as the 
smallest characteristic scale of the structures which develop in the flow.
Typically, we find that the saturation scale $l_S$ is of order of 
$2\div 3$ times the relevant Kolmogorov scale $l_d$, showing that 
the extent of structures in their smallest dimension is controlled by
dissipative effects.
On the other hand, power-law scaling sets in at the scale $l_{\star}$ which is 
found to be of the same order as the Taylor microscale $\lambda$. 
This point can be understood by a simple geometrical argument 
developed in the next section. 
All these results obtained for the current are confirmed for the other fields 
$\omega$, $\omega^+$ and $\omega^-$. 
The values of the cancellation exponents and those of the typical scales 
related to the partition function for all fields are reported in Table 
\ref{table}. 
It can be observed that the rotationals of the Els\"asser fields both present 
the same behavior, while different values of the parameters are 
found for the vorticity and the current. Note also that
the range of fit for the vorticity is shorter than those for 
the others fields (not shown). 

The higher value of $\kappa^{\omega}$ 
reveals that the presence of structures inhibit 
the cancellations in a less efficient way for the vorticity than for the other 
fields. Conversely, the current structures inhibit cancellations more 
efficiently than all other fields.

\section{Cancellation exponents and dimensions of structures}
\label{sec5}

It is clear that cancellation exponents characterize in some way the geometry 
of structures, as exemplified by the fact that a smooth (continuous) field
will give a zero cancellation exponent, while a field made of uncorrelated
point-like objects will have a cancellation exponent~$\kappa=d/2$. In order to
quantify the transition between these two limits, we will use a simple
geometrical model that we now introduce.

The relevant Taylor scale~$\lambda$ gives us a mean scale over which
a field is correlated. However, it does not tell us the geometry of the
structures present at that scale.  These could be correlated in some
directions for scales much larger than~$\lambda$, and/or already
uncorrelated in some other directions.  Let us now assume that the field
is smooth (correlated) in $D$~dimensions with a cutoff scale of~$\lambda$,
and uncorrelated in the other $d-D$~dimensions.  Of course, if the field
becomes completely regular below some scale~$l_S$, the corresponding
dimension at that scale will be~$D=d$.  On the other hand, if structures
have some largest extent~$l_u$, the field will consist of isolated, and
thus nearly uncorrelated, objects, showing an apparent dimension~$D=0$.
In the intermediate region, over which the topology of the structures will
have an effect, the partition function of the current~$J$ for example, or of
any component of a field in dimension $d$, can be written as
 \begin{eqnarray}
 \chi(l) &=& \sum_{Q_i(l)}\left|{\int_{Q_i(l)}d\gra{r}\,J(\gra{r}) \over
             \int_{Q(L)}d\gra{r} |J(\gra{r})|}\right| \nonumber \\
         &\sim& {1 \over L^d J_{\rm rms}} \left({L \over l}\right)^d
                |\int_{Q(l)}d\gra{r}\,J(\gra{r})|
 \label{integrali}               
 \end{eqnarray}
where homogeneity is used to replace the sum over all subsets~$Q_i(l)$
by $(L/l)^d$ times the integral over a generic box $Q(l)$ of size $l$. 
Moreover, we have approximated the field absolute value integral by its typical 
value~$L^d J_{\rm rms}$.  Indeed,
because of the absolute values, no cancellations can occur in the denominator.
Now, the integral on the subset~$Q(l)$ can be done by integrating over regular
domains of size $\lambda^d$ and splitting the number of contributions between 
the correlated dimensions and the uncorrelated ones.  
The smooth dimensions will give a contribution proportional to their 
area~$(l/\lambda)^D$.  
The uncorrelated dimensions will behave like the integral of an uncorrelated 
field, giving a value proportional to the square root of their area, that
is~$(l/\lambda)^{(d-D)/2}$.  Collecting all contributions, we see that
$\chi(l)$ will behave as
 \begin{eqnarray}
 \chi(l) &\sim& {\lambda^d J_{\rm rms} \over L^d J_{\rm rms}}
                \left({L \over l}\right)^d \left({l \over \lambda}\right)^D
                \left({l \over \lambda}\right)^{d-D \over 2} \nonumber \\
         &\sim& \left({l \over \lambda}\right)^{-{d-D \over 2}} \sim
                \left({l \over \lambda}\right)^{-\kappa} \ .
 \label{fractal}
 \end{eqnarray}
The cancellation exponent can thus be interpreted as half the codimension
of the flow structures, $\kappa=(d-D)/2$.  
Note that this interpretation remains valid at small
scales where~$\kappa=0$ when~$D=d$ and at scales larger or similar to the
integral scale where~$\kappa=d/2$ when~$D=0$.

Specializing to our case with $d=2$, the value of~$\kappa^\pm=0.5$
that we found indicates that structures are similar to
filaments~$D^\pm=1.0\pm0.12$ for the $\gra{\omega^\pm}$ fields.
This result for the Els\"asser variables rotationals extends the previous
observation of sheet-like structures for the current 
in 3D MHD~\cite{politano95,kerrmhd,bisk3d}, having
a corresponding signature of filaments in 2D cuts.  

The current displays a slightly smaller cancellation
exponent, that leads to a dimension of~$1.14\pm0.12$, which can be interpreted
by observing that the current structures are slightly fatter than filaments
and show some thickness, likely due to reconnection taking place within 
current sheets.

The vorticity field has a significantly larger cancellation exponent, that would
give a dimension for structures~$D^\omega=0.62\pm0.24$, smaller than the one of
a filament structure. In fact, a close inspection in physical
space shows that vorticity structures, although globally thin and elongated, 
are more complex than those of the current field (see plots at $t=6.3\sim
26\,\tau_{NL}$, and at $t=6.93\sim 39\,\tau_{NL}$ in Figure~\ref{fig5}). 
For example, as already known~\cite{politano90,matthaeus}, 
quadrupole vorticity structures are associated with quasi-linear current 
sheets in the simplest reconnection configuration. 
This could explain the smaller dimension found for the vorticity structures 
characterizing what can be seen as a ``disrupted filament''.

\section{Conclusion and discussion}
\label{sec6}

In order to characterize the oscillating behavior of the dynamical structures
present in conductive turbulent flows, we perform a signed measure analysis
of data obtained from direct numerical integration of the two-dimensional MHD
equations. We obtain that the measure is sign singular, with a clear scaling
range for the cancellations between the structures, either for the current and
the vorticity field, or for the rotational of the Els\"asser variables. The
cancellations are inhibited for scales smaller than the Taylor scales while
they stop near the dissipative scales. 
Note that the correspondence -- as displayed in Table~\ref{table} -- 
between the typical scales for cancellation and 
those of the turbulent fluid are obtained in the 
framework of MHD flows presenting a weak correlation between the velocity and 
the magnetic fields and for a magnetic Prandtl number of unity.

Moreover, by a simple geometrical argument, we link the
cancellation exponent evaluated here to the fractal dimension of the
structures in the flow. This gives us informations about the topology of
the structures. This interpretation can in fact also be extended. 
Indeed, the partition function is only a first-order moment, and is thus 
unable to distinguish, for example, between a regular behavior over some 
fractal set, and a self-similar behavior over the whole space, as well as 
intermediate situations.  The distinction could only be made by looking at 
higher order moments~\cite{vains2}.  This point is left for future study.  
But we can already generalize the geometrical argument given in the previous 
section to the case where the field is not locally smooth, by introducing a
single local scaling exponent~$h$ such that 
(specializing to the magnetic field) $\|\Delta J(s)\|\sim s^{h-1}$ 
on the structures of dimension~$D$. Note that, because of the presence of
structures, $h$ is {\em not\/} the exponent of the first-order structure
function of~$J$, the latter containing contributions both of structures and
background. In this picture, the last integral in expression~(\ref{integrali}) 
becomes
\begin{eqnarray}
\nonumber
 \int_{Q(l)}\!\!\!\!d^{\raisebox{2pt}{\scriptsize{\it d}}}\!\!s\,J(\gra{s}) & = & 
 J_{\rm rms}\;\int_{Q(l)}\!\!\!\!d^{\raisebox{2pt}{\scriptsize{\it D}}}\!\!\!s\,
                                \left({s \over \lambda}\right)^{h-1}
            \;\int_{Q(l)}\!\!\!\!d^{\raisebox{2pt}{\scriptsize{\it d-D}}}\!\!\!\!\!\!\!s  \\
\nonumber
& = & J_{\rm rms}\lambda^D\,\left({l \over \lambda}\right)^{D+h-1}\,
                 \lambda^{d-D}\,\left({l \over \lambda}\right)^{d-D \over 2}\\
\nonumber
  & = & \lambda^d J_{\rm rms}\,\left({l \over \lambda}\right)^{D+h-1}\,
                               \left({l \over \lambda}\right)^{d-D \over 2} 
\end{eqnarray}
giving a cancellation exponent $\kappa = (d-D)/2 + (1-h)$. 
In general, as we already pointed out, it is impossible to separate the
contributions of the two terms, unless one makes further hypotheses.

Let us consider the particular case of a space filling object ($D=d$) which
would not be smooth at small scales, that is with a H\"older self-similarity
exponent~$h$ smaller than one. The cancellation exponent of its rotational
would then be simply~$\kappa=1-h$. 
Note that if the object is smooth, with $h=1$, this is consistent with
$\kappa=0$ obtained from (\ref{fractal}). On the other hand, if the field
values are completely independent, we find $\kappa=1$, that can be easily
checked by direct numerical simulations~\cite{consolini}.
Using this interpretation, in the case of the velocity rotational that has a
cancellation exponent $\kappa=0.69 \pm 0.12$, we are 
led to infer an H\"older exponent for the velocity~$h=0.31\pm0.12$, 
close to the Kolmogorov value~$1/3$ for the velocity field of a neutral
turbulent flow.
This argument is clearly incomplete because of the presence of structures in
the fields, which are even more conspicuous for the rotationals of the magnetic
and Els\"asser fields. 
Higher resolution simulations allowing the computation of the full spectrum of 
higher order cancellation exponents will be necessary to disentangle the 
contributions of geometry and differentiability to the intermittency 
of MHD turbulence.

\acknowledgments
The numerical simulations were performed at IDRIS (Orsay). We received partial
financial support from the CNRS Program PNST.

\newpage

\begin{table}
\caption[Table]{Estimated values for the typical scales of the flow (see text 
for definitions), the
cancellation exponents and the fractal dimensions of the structures 
for the four fields, averaged on ten temporal snapshots; $l_d$ is the 
dissipation scale, $\lambda$ the Taylor scale, $l_*$ denotes the end of the 
inertial range and $l_S$ is the saturation scale.}
\begin{center}
\begin{tabular}{ccccccc}
Field&$l_d/L\times10^{3}$&$\lambda/L\times10^{3}$&$l_S/L\times10^{3}$&
  $l_{\star}/L\times10^{3}$&$\kappa$&$D$\\
\hline&&&&&&\\[-8pt]
$   J$     &      ---      &      ---      & $5.0 \pm 0.8$ & $15$ 
   & $0.43 \pm 0.06$ & $1.14 \pm 0.12$\\
$ \omega$  &      ---      &      ---      & $6.9 \pm 1.0$ & $20$ 
   & $0.69 \pm 0.12$ & $0.62 \pm 0.24$\\ 
$\omega^+$ & $2.1 \pm 0.7$ & $17 \pm 9$ & $5.4 \pm 1.0$ & $15$ 
   & $0.50 \pm 0.06$ & $1.00 \pm 0.12$\\
$\omega^-$ & $2.4 \pm 0.7$ & $18 \pm 8$ & $5.4 \pm 1.0$ & $15$ 
   & $0.50 \pm 0.07$ & $1.00 \pm 0.12$\\
[2pt]\end{tabular}
\label{table}
\end{center}
\end{table}
\vskip 3cm

\newpage

  \bfigc
  \caption{A snapshot of the current~$J$ at time~$t=7.3$ in grey levels (top), 
  and a pseudo-3D perspective view of the same field (bottom) showing the
  existence of structures of different signs at all scales.}
  \label{fig1} 
  \efigc

\newpage

  \bfigc
\epsfxsize=7.2cm
\epsffile{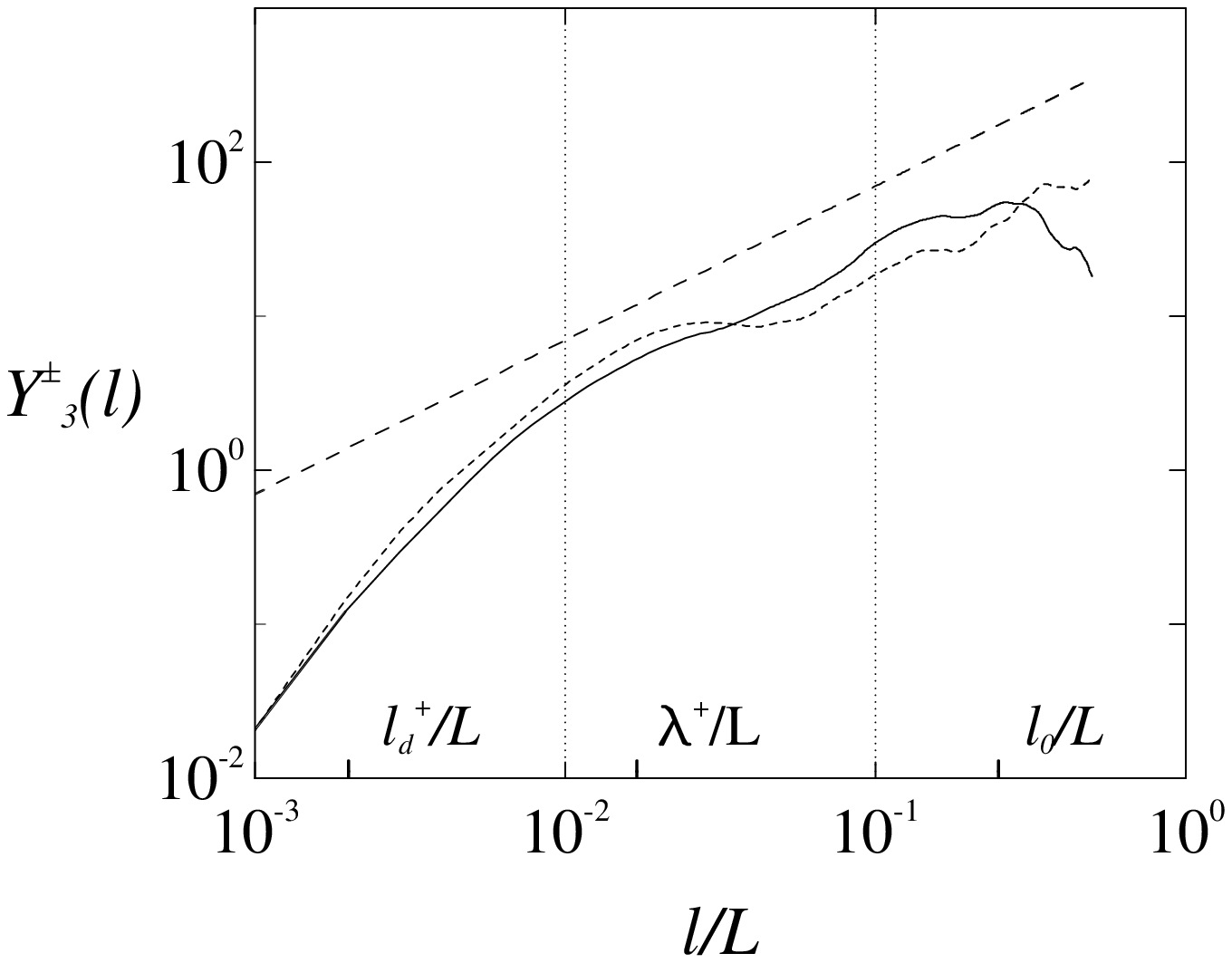}
\caption{The time averaged third order correlators~$Y_3^+(l)$ (continuous line)
and~$Y_3^-(l)$ (dashed line) in a log-log plot. The straight line is a simple 
linear law, and is displayed as reference. 
The dissipation scale~$l_d^+$, the Taylor microscale~$\lambda^+$ 
and the integral scale~$l_0$ are indicated, as well as the bounds of the 
inertial range.}     
\label{fig2} 
  \efigc

\newpage

  \bfigc
  \caption{The coarse-grained signed measure of the current~$J$ at time~$t=7.3$ 
  for four different box sizes, namely $l/L=0.001$, $l/L=0.016$, $l/L=0.059$, 
  $l/L=0.12$, from top to bottom. Colors range from cyan for negative~$J$ values
  to yellow for positive ones, going through blue and brown. Cancellations at
  large scales are responsible for the decrease in magnitude of the measure.}     
  \label{fig3} 
  \efigc

\newpage

  \bfigc
\epsfxsize=7.2cm
\epsffile{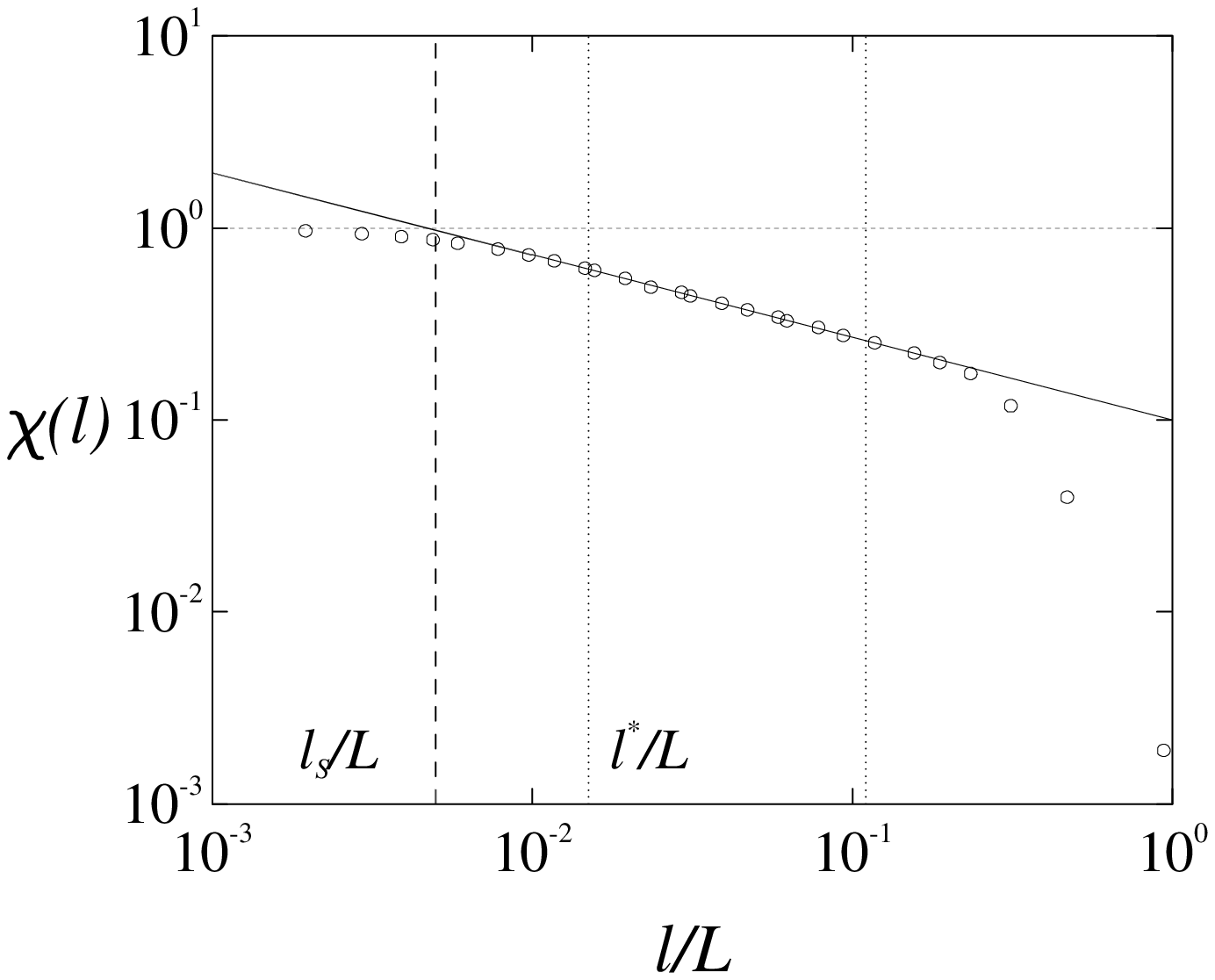}
\caption{For the current~$J$, the log-log plot of~$\chi(l)$ (averaged over 
ten snapshots) {\it versus} the subsets size~$l/L$. 
The solid line represents the best fit with the power-law (\ref{powlaw}). 
The constant value $\chi_S(l)=1$ is plotted (the horizontal 
dashed line), and the saturation scale $l_S/L=0.005$ is indicated by the 
vertical dashed line on the left. 
The vertical dotted lines indicate the range of the fit, which lies between 
$l_{\star}/L=0.015$ and $l_1/L=0.12$.}     
\label{fig4} 
  \efigc

\newpage

  \bfigc
  \caption{Examples of contour lines of vorticity (left panel) and 
  current (right panel) structures, at the time $t=6.3$ (top) and $t=6.93$
  (bottom). The axes indicate the location of the snapshots within the
  $2\pi$-periodic box.}     
  \label{fig5}
  \efigc

\end{document}